\documentclass{iau} 
\usepackage[T1]{fontenc}
\usepackage{graphicx}
\usepackage{natbib}
\usepackage{amssymb}
\usepackage{mathabx}

\newcommand{\msun}{${\mathrm{M}}_{\odot}$}

\title[Planets Transiting White Dwarfs with ZTF] 
{The Search for Planet and Planetesimal Transits of White Dwarfs with the Zwicky Transient Facility}

\author[]   
{Keaton J.\ Bell$^{1}$}

\affiliation{$^1$NSF Astronomy and Astrophysics Postdoctoral Fellow and DiRAC Institute Fellow, \\ Department of Astronomy, University of Washington, Seattle, WA 98195, USA \\ email: {\tt keatonb@uw.edu} }

\pubyear{2019}
\volume{357}  
\jname{White Dwarfs as Probes of Fundamental Physics and Tracers of Planetary, Stellar, and Galactic Evolution}
\begin{document}

\maketitle

\begin{abstract}
Planetary materials orbiting white dwarf stars reveal the ultimate fate of the planets of the Solar System and all known transiting exoplanets. Observed metal pollution and infrared excesses from debris disks support that planetary systems or their remnants are common around white dwarf stars; however, these planets are difficult to detect since a very high orbital inclination angle is required for a small white dwarf to be transited, and these transits have very short (minute) durations.  The low odds of catching individual transits could be overcome by a sufficiently wide and fast photometric survey.  I demonstrate that, by obtaining over 100 million images of white dwarf stars with 30-second exposures in its first three years, the Zwicky Transient Facility (ZTF) is likely to record the first exoplanetary transits of white dwarfs, as well as new systems of transiting, disintegrating planetesimals. In these proceedings, I describe my project strategy to discover these systems using the ZTF data.
\keywords{planetary systems, white dwarfs, surveys}
\end{abstract}

\firstsection 
\section{Introduction}

White dwarf stars (WDs) represent the final evolutionary stage of >97\% of stars \citep[e.g.,][]{Williams2009}, including the host stars of every known transiting exoplanet. The present population of planets orbiting WDs reveals the fates of these planets that missions like \emph{Kepler} have shown to be ubiquitous around main sequence stars \citep[e.g.,][]{Hsu2019}.  By virtue of its high volumetric survey speed \citep{Bellm2016}, the Zwicky Transient Facility (ZTF) holds the promise of detecting the first members of this important exoplanet population, the details of which will inform our theories of planetary migration and survival in post-main-sequence circumstellar environments.

Despite the fact that no intact planets orbiting WDs have been previously identified, we have compelling observational evidence that such planets are common. The high gravities of WDs cause metals to diffuse quickly below their photospheres \citep{Koester2006}, yet 27--50\% of WDs are metal polluted \citep{Koester2014}, implying that accretion of metals is ongoing. Chemical abundance analyses support that these metals come from rocky planetary debris \citep[e.g.,][]{Gaensicke2012}. Furthermore, debris disks are detected via infrared excesses from $\approx2\%$ of these systems \citep{Rebassa-Mansergas2019}. Recently, disintegrating planetesimals have even been discovered to orbit a couple of specific metal-polluted WDs \citep{Vanderburg2015,Manser2019}. Some of my own high speed photometry from McDonald Observatory of the transiting debris system WD\,1145+017 is displayed in Figure~\ref{fig:wd1145}.

\begin{figure}
\centering
\includegraphics[width=0.6\textwidth]{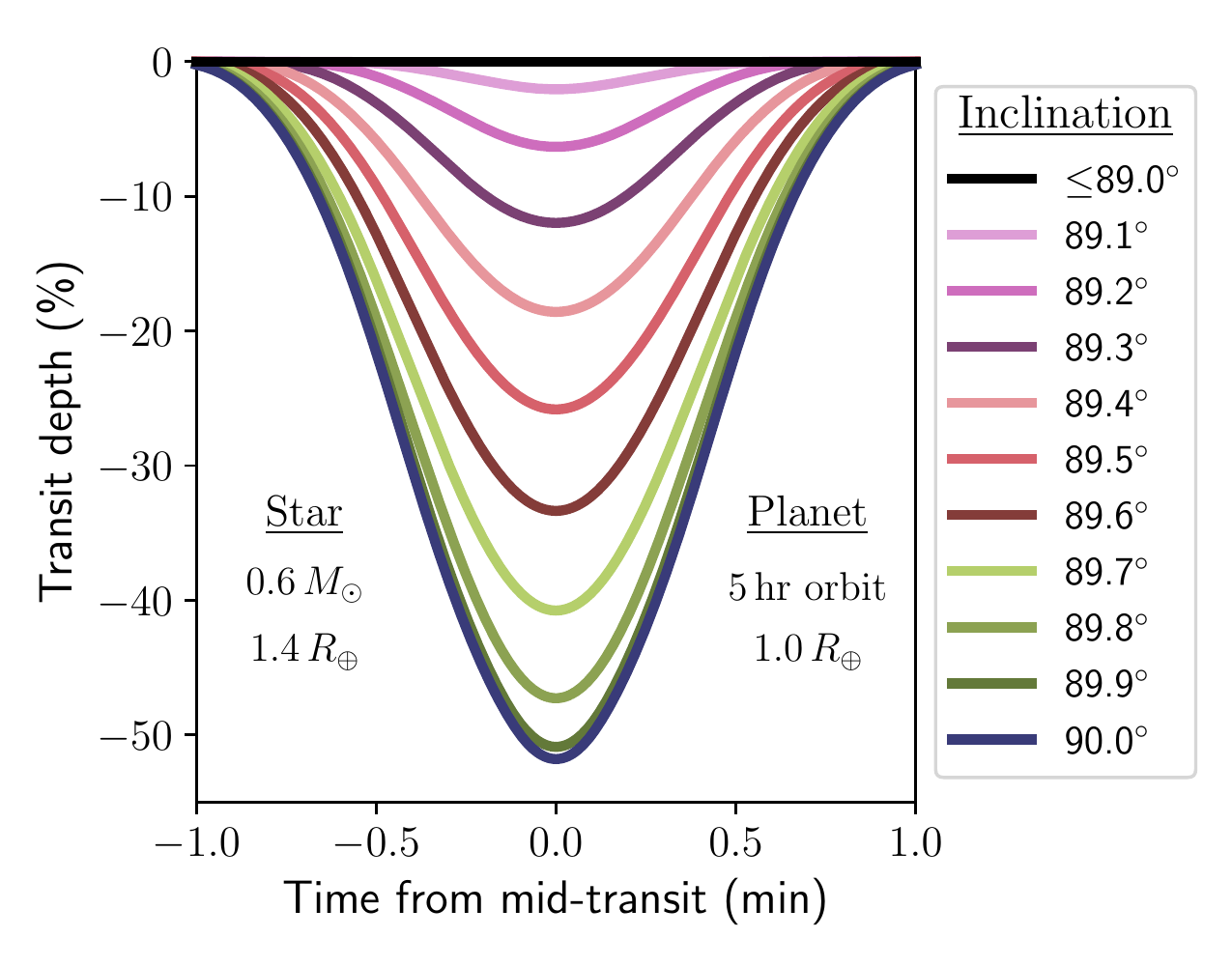}
\caption{Model light curves of a 1\,$R_\Earth$ planet transiting a WD at different inclination angles, simulated with the {\sc batman} package \citep{Kreidberg2015} accounting for 30-second ZTF exposures.}\label{fig:lims}
\end{figure}

Detecting planets around WDs is observationally very challenging since transits of these roughly Earth-sized stars have short ($\sim$\,minute) durations and require nearly edge-on alignments. Still, transits of WDs can be very deep (even occulting), producing unmistakable signals in well-timed images.  Figure~\ref{fig:lims} explores the transit profile of an Earth-sized planet in a 5-hour orbit around a typical 0.6\,\msun\ WD at very high inclination angles.

Previous searches for such dips in samples of $\sim$1000 WDs have yielded no transit detections (from, e.g., WASP, \citealt{Faedi2011}; Pan-STARRS, \citealt{Fulton2014}; \emph{K2}, \citealt{vanSluijs2018}; DECam, \citealt{Dame2019}; GALEX, \citealt{Rowan2019}). Discovering the first planets to transit WDs will require orders of magnitude larger sample sizes. \citet{Cortes2019} and \citet{Lund2018} have simulated the expected yield for WD transits after 10 years of Large Synoptic Survey Telescope (LSST) operations (ca.\ 2033), anticipating $\sim$1000 detections.  Because LSST is a deep survey, opportunities to follow-up candidate WD transits with large telescopes will be extremely limited.  The magnitude range of ZTF, on the other hand, is readily accessible to more modest follow-up facilities, and I demonstrate in these proceedings that the current potential for discovering WD transits with the modern ZTF survey has been largely overlooked.

\section{Detectability of Planet Transits with ZTF}

\begin{figure}
\centering
\includegraphics[width=1\textwidth]{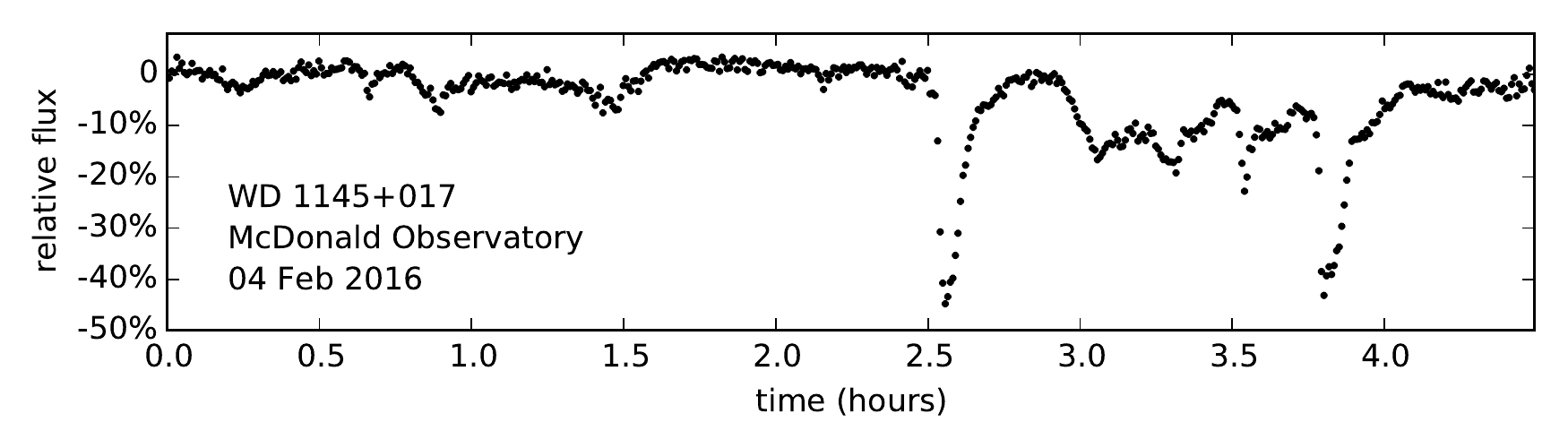}
\caption{WD\,1145+017 light curve from McDonald Observatory covering one orbital period.}\label{fig:wd1145}
\end{figure}

\begin{figure}
\centering
\includegraphics[width=0.6\textwidth]{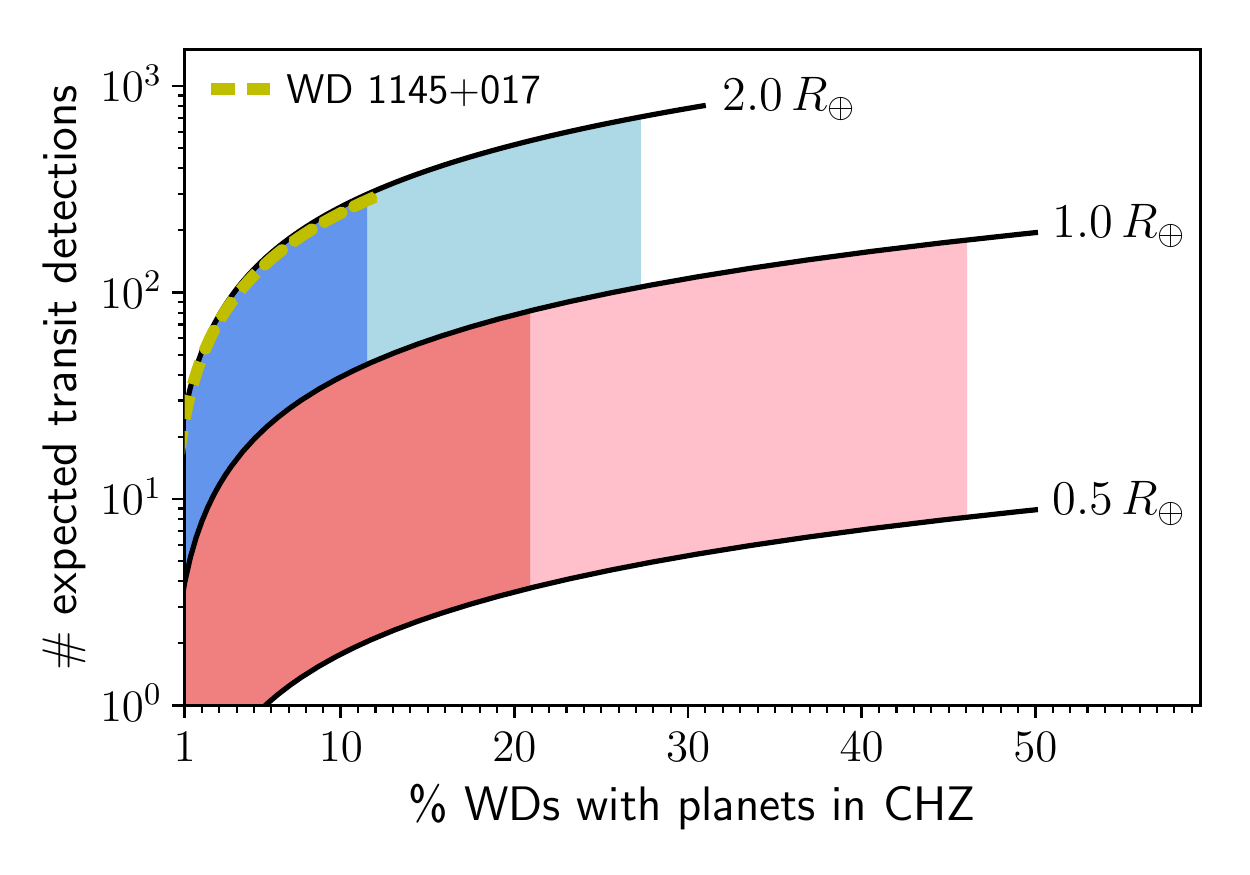}
\caption{Simulated number of transit detections expected from the three-year ZTF survey for different percentages of WDs hosting 0.5, 1, and 2\,$R_\Earth$ planets in the continuously habitable zone. Regions are shaded below the 95\% (light) and 68\% (dark) confidence upper limits on planet occurrence from \citet{vanSluijs2018}. The dashed line shows the expected detections of disintegrating planetesimals transiting WDs based on the first such system, WD\,1145+017.}\label{fig:simresults}
\end{figure}

Thanks to \emph{Gaia} astrometry, we now know the identities \citep{GentileFusillo2018} of most of the $\approx$220{,}000 WDs in the ZTF survey footprint (declination > $-31\,\deg$) brighter than the median $5\sigma$ survey depth of $g=20.8$\,mag, each of which will receive a median $\approx870$ observations over three years \citep{Bellm2014}.
I use the {\sc batman} package \citep{Kreidberg2015} to synthesize transit light curves for planets that uniformly and isotropically populate the period range of the continuously habitable zones \citep[4.5\,hr\,$<P_\mathrm{orb}<32$\,hr;][]{Agol2011} around typical 0.6\,\msun\ WDs \citep[e.g.,][]{Tremblay2013}.
A transit detection is considered significant when the measured flux decrease exceeds a $5\sigma$ threshold, where $\sigma$ is the photometric error achieved through the magnitude range observed by ZTF \citep{Bellm2019}.

Figure~\ref{fig:simresults} displays the expected WD planet yield from ZTF. The solid lines show the number of $5\sigma$ transit detections for a given fraction of WDs that host planets of radii 0.5, 1, and 2 $R_\Earth$ in their continuously habitable zones.  The shaded regions are consistent with recent constraints based on $\sim$1000 non-detections from \emph{K2} \citep{vanSluijs2018}. 
If as few as 1\% of WDs host 1--2\,\emph{R}$_\Earth$ planets in their continuously habitable zones, ZTF will detect $\gtrsim$\,10 exoplanet transits during its 3-year survey.

ZTF is also expected to discover many new examples of disintegrating planetesimal systems transiting WDs like WD\,1145+017, which was found in \emph{K2} data \citep{Vanderburg2015}.  Using the ground-based light curve of this system obtained with 30-second exposures from the McDonald Observatory 2.1-meter Otto Struve Telescope (Figure~\ref{fig:wd1145}) as a template, I include estimates of the number of similar systems that ZTF will detect with the dashed line in Figure~\ref{fig:simresults}. In fact, the second such transiting planetesimal system has already been discovered from the first public release of ZTF data \citep[ZTF\,J013906.17+524536.89;][]{Vanderbosch2019}.

\section{Project Strategy}

This work is the topic of the project ``Discovering the First Exoplanets Around White Dwarf Stars with the Zwicky Transient Facility'' that was granted a three-year NSF Astronomy and Astrophysics Postdoctoral Fellowship award \#1903828 (PI: Bell). The project is designed with the following strategy:
\begin{enumerate}
    \item Use the Astronomy eXtensions for Spark \citep[AXS;][]{Zecevic2019} tools to efficiently identify candidate transits of likely WDs in the large ZTF data set (including both public and proprietary partnership data).
    \item After validating the data quality of ZTF measurements that are consistent with significant brightness decreases, obtain follow-up high-speed photometry from Apache Point Observatory (APO), confirming the transits and resolving their shapes and periods.
    \item Obtain spectra of the WD hosts at APO to characterize stellar atmospheric parameters ($T_\mathrm{eff},\log{g}$) and to rule out binarity for targets that exhibit light curves consistent with binary eclipses.
    \item Begin long-term monitoring of short-period transits to measure potential timing variations that probe the population of more distant planets that are unlikely to transit.
    \item Also identify signatures of transits of WDs by extended planetesimal debris (such as WD\,1145+017, \citealt{Vanderburg2015}; ZTF\,J013906.17+524536.89, \citealt{Vanderbosch2019}), and begin long-term photometric monitoring for dynamical variations \citep[e.g.,][]{Gaensicke2016}.
    \item At the conclusion of the three-year ZTF survey, use the total number of transit detections and non-detections to place strict constraints on the occurrence rates of planets of different sizes and orbital periods around WDs.
\end{enumerate}

\acknowledgements This material is based upon work supported by the National Science Foundation under Award No.\ 1903828.

\end{document}